\documentclass[10pt,journal]{IEEEtran}
\newlength{\saveparindent}
\newlength{\saveparskip}
\newcount\proofqeded
\newcount\proofended
\def\qed{ {\hspace{5pt}\rule[-1pt]{3pt}{9pt}}
\end{rm}\addtolength{\parskip}{-0pt}
\setlength{\parindent}{\saveparindent}
\global\advance\proofqeded by 1 }

{\proofstart}%
{\ifnum\proofqeded=\proofended\qed\fi \global\advance\proofended by 1
\medskip}
\makeatletter
\def\proofstart{\@ifnextchar[{\@oprf}{\@nprf}}
\def\@oprf[#1]{\begin{rm}\protect\vspace{6pt}\noindent{\bf Proof of #1:\
}%
\addtolength{\parskip}{5pt}\setlength{\parindent}{0pt}}
\def\@nprf{\begin{rm}\protect\vspace{6pt}\noindent{\bf Proof:\ }%
\addtolength{\parskip}{5pt}\setlength{\parindent}{0pt}}

\newcounter{ctr}

\newcounter{ectr}

\newlength{\savejot}
\newcommand{\concat}{\:\|\:}

\newcommand{\calF}{{\cal F}}

\makeatletter

\usepackage{float}
\usepackage{booktabs} 
\usepackage{import}
\usepackage{graphicx}
\usepackage{subcaption}
\usepackage{caption}
\usepackage[utf8]{inputenc}
\usepackage[export]{adjustbox}
\usepackage{wrapfig}
\usepackage{amsmath}
\usepackage{algorithm}
\usepackage[noend]{algpseudocode}
\usepackage{amssymb}
\usepackage{multirow}
\usepackage{enumitem}
\usepackage{varwidth}
\usepackage{tabularx}
\usepackage{xcolor}

\usepackage{balance}

\usepackage{multicol}

\ifCLASSINFOpdf
\else
\fi
\begin{document}
%
\title{Lightweight Encryption and Anonymous Routing in NoC based SoCs}
%
%
%

\author{Subodha~Charles,~\IEEEmembership{Member,~IEEE,}
        and~Prabhat~Mishra,~\IEEEmembership{Fellow,~IEEE}
\thanks{An early version of this work has appeared as a non-reviewed  book chapter for the general audience~\cite{charles2021lightweight}.}
\thanks{S. Charles is with the Department of Electronic and Telecommunication Engineering, University of Moratuwa, Colombo, Sri Lanka. e-mail: scharles@uom.lk.}
\thanks{P. Mishra is with the Department
of Computer \& Information Science \& Engineering, University of Florida, Gainesville, Florida, USA. e-mail: prabhat@ufl.edu.}
}

\maketitle

\begin{abstract}
Advances in manufacturing technologies have enabled System-on-Chip (SoC) designers to integrate an increasing number of cores on a single SoC. Increasing SoC complexity coupled with tight time-to-market deadlines has led to increased utilization of Intellectual Property (IP) cores from third-party vendors. SoC supply chain is widely acknowledged as a major source of security vulnerabilities. Potentially malicious third-party IPs integrated on the same Network-on-Chip (NoC) with the trusted components can lead to security and trust concerns. While secure communication is a well-studied problem in the computer networks domain, it is not feasible to implement those solutions on resource-constrained SoCs. In this paper, we present a lightweight encryption and anonymous routing protocol for communication between IP cores in NoC based SoCs. Our method eliminates the major overhead associated with traditional encryption and anonymous routing protocols using a novel secret sharing mechanism while ensuring that the desired security goals are met. Experimental results demonstrate that existing security solutions on NoC can introduce significant (1.5X) performance degradation, whereas our approach provides the same security features with minor (4\%) impact on performance.
\end{abstract}

\begin{IEEEkeywords}
Encryption, Anonymous Routing, Network-on-Chip, System-on-Chip, Hardware Security
\end{IEEEkeywords}

%
\IEEEpeerreviewmaketitle

\section{Introduction} \label{sec:introduction}

\IEEEPARstart{T}{he} growth of general purpose as well as embedded computing devices has been remarkable over the past decade. This was mainly enabled by the advances in manufacturing technologies that allowed the integration of many heterogeneous components on a single System-on-Chip (SoC). The tight time-to-market deadlines and increasing complexity of modern SoCs have led manufacturers to outsource intellectual property (IP) cores from potentially untrusted third-party vendors~\cite{huang2018scalable,farahmandi2019system}. Therefore, the trusted computing base of the SoC should exclude the third-party IPs. In fact, measures should be taken since malicious third-party IPs (M3PIP) can launch passive as well as active attacks on the SoC~\cite{mishra2017hardware}. Such attacks are possible primarily because the on-chip interconnection network that connects SoC components together, popularly known as Network-on-Chip (NoC), has visibility of the entire SoC and the communications between IP cores. 
Previous efforts have developed countermeasures against stealing information~\cite{sepulveda2017towards}, snooping attacks~\cite{chittamuru2018soteria}, and even causing performance degradation by launching denial-of-service (DoS) attacks~\cite{charles2019real}. In this paper, we present a countermeasure for M3PIPs operating under the following architecture and threat models.

\begin{figure}[h]
    \centering
    \includegraphics[width=0.98\columnwidth]{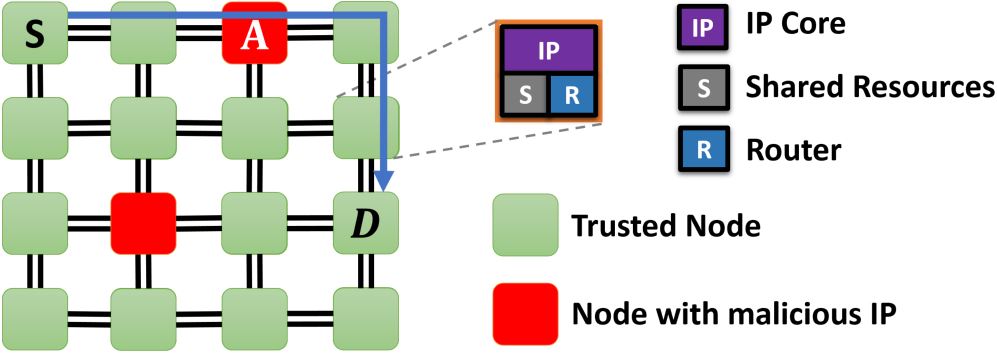}
    \caption{Overview of a typical SoC architecture with IPs integrated on a Mesh NoC.}
    \label{fig:trust-architecture} 
\end{figure}

\textbf{Threat Model: } Figure~\ref{fig:trust-architecture} shows an SoC with heterogeneous IPs integrated on a Mesh NoC. The two nodes marked as $S$ (source) and $D$ (destination) are trusted IPs communicating with each other. M3PIPs integrated on the SoC (nodes shown in red) have the following capabilities when packets pass through their routers:
\begin{itemize}
    \item They can steal information if data is sent as plaintext.
    \item If data is encrypted and header information is kept as plaintext, they can gather packets generated from the same source and intended to the same destination and launch complex attacks such as linear/differential cryptanalysis since they belong to the same communication session.
    \item When multiple M3PIPs are present on the same NoC, they can share information and trace messages.
    \item An M3PIP can compromise the router attached to it and gather information stored in the router. This can leak routing information. Assuming only some of the IPs are acquired from untrusted third-party vendors, all routers will never be compromised at the same time.
\end{itemize}

It is not feasible to utilize traditional security methods (encryption, authentication,  etc.) in resource-constrained embedded devices. Previous studies explored lightweight security architectures to mitigate threats. Previous work on lightweight encryption proposed smaller block and key sizes, less rounds of encryption and other hardware optimizations~\cite{kapoor2013security}. Irrespective of the optimizations, these methods still have complex computations that take several cycles. In this paper, we propose a \textit{Lightweight Encryption and Anonymous Routing protocol for NoCs} (LEARN) that requires only few addition and multiplication operations for encryption. We are able to eliminate the traditional encryption methods consisting of ciphers and keys entirely by using the \textit{secret sharing} approach proposed by Shamir~\cite{shamir1979share} without compromising the security guarantees. Furthermore, our framework supports anonymous routing such that an intermediate node can neither detect the origin nor the destination of a packet. Major contributions of this paper can be summarized as follows:
\begin{itemize}
    \item We propose an anonymous routing scheme that hides both source and destination information making the packets untraceable. Launching attacks on encrypted data passing through a given router becomes more difficult when the packets are untraceable and origins are unknown.
    \item We develop a lightweight encryption scheme that is based on secret sharing.
    \item We demonstrate that our approach is lightweight compared to existing encryption methods as well as traditional anonymous routing methods such as onion routing.
\end{itemize}

The remainder of the paper is organized as follows. Section~\ref{sec:background} introduces some concepts used in this paper and presents related efforts. Section~\ref{sec:motivation} motivates the need for our work. Section~\ref{sec:learn} describes our lightweight encryption and anonymous routing protocol.
Section~\ref{sec:results} presents the experimental results. 
Section~\ref{sec:discussion} discusses possible further enhancements to our approach.
Finally, Section~\ref{sec:conclusion} concludes the paper.


\section{Background and Related Work} \label{sec:background}

This section introduces some of the key concepts used in our proposed framework. The first three sections introduce symmetric and asymmetric encryption, Lagrangian polynomial based interpolation and anonymous routing. The last section discusses prior work in lightweight encryption and anonymous routing to explain how our proposed approach differs from them.

\subsection{Symmetric and Asymmetric Encryption} \label{ssec:intro_crypto}

{\it Symmetric Encryption: } 
A symmetric encryption scheme takes the same key $K$ for both decryption and encryption. The encryption algorithm $E$ produces the \textit{ciphertext} $C$ by taking the key $K$ and a \textit{plaintext} $M$ as inputs, This is denoted by $C \leftarrow E_K(M)$. Similarly, the \textit{decryption} algorithm $D$ denoted by $M \leftarrow D_K(C)$, takes a key $K$ and a ciphertext $C$ and returns the corresponding $M$. The correctness of the scheme is confirmed when any sequence of messages $M_1,..., M_u$ encrypted under a given key $K$ produces $C_1 \leftarrow E_K(M_1)$, $C_2 \leftarrow E_K(M_2)$,..., $C_u \leftarrow E_K(M_u)$, and is related as $D_K(C_i) = M_i$ for each $C_i$.

{\it Asymmetric Encryption: } 
In asymmetric encryption, also known as public key encryption, different keys are used for encryption and decryption. Encryption is done using the \textit{public key} that is publicly known by all the entities in the environment. An entity $B$ that wants to send a message $M$ to another entity $A$  will encrypt the message using $A$'s public key (with public key $PK_A$) to produce ciphertext $C$ denoted by $C \leftarrow E_{PK_A}(M)$. The ciphertext can only be decrypted by $A$'s \textit{secret key (private key)} $SK_A$ corresponding to $PK_A$. $SK_A$ is known by only $A$, and therefore, only $A$ can decrypt $C$ to produce $M$ denoted by $M \leftarrow D_{SK_A}(C)$.

\subsection{Secret Sharing with Polynomial Interpolation} \label{ssec:intro_ss}

Shamir's secret sharing \cite{shamir1979share} is based on a property of \textit{Lagrange polynomials} known as the $(k,n)$ threshold. It specifies that a certain secret $M$ can be broken into $n$ parts and $M$ can only be recovered if at least $k$ $(k \leq n)$ parts are retrieved. The knowledge of less than $k$ parts leave $M$ completely unknown. Lagrange polynomials meet this property with $k = n$. A Lagrange polynomial is comprised of some $k$ points $(x_0,y_0)$, ..., $ (x_{k-1},y_{k-1})$ where $x_i \neq x_j$ $(0 \leq i, j \leq k-1)$. A unique polynomial of degree $k-1$ can be calculated from these points:
\begin{equation} \label{eq:lag_sum}
    L(x) = \sum_{j=0}^{k-1} l_j(x) \cdot y_j,
\end{equation}
where
\begin{equation} \label{eq:lag_coef}
    l_j(x) = \prod_{i=0, i \neq j}^{k-1} \frac{x-x_i}{x_j - x_i}
\end{equation}
Any attempt to reconstruct the polynomial with less than $k$ or incorrect points will give the incorrect polynomial with the wrong coefficients and/or wrong degree.

$L(x)$ forms the interpolated Lagrange polynomial, and $l_j(x)$ is the Lagrange basis polynomial. In order to a share a secret using this method, a random polynomial of degree $k-1$ is chosen. It takes the form of $L(x) = a_0 + a_1x + a_2x^2 + ... + a_{k-1}x^{k-1}$. The shared secret $M$ should be set as $a_0=M$, and all the other coefficients are chosen randomly. Then a simple calculation at $x = 0$ would yield the secret ($M = L(0)$). In this case, $k$ points on the curve are chosen at random and distributed together with their respective $l_j(0)$ values - the Lagrangian coefficients. To retrieve $M$, all the parties should share their portions of the secrets. Once all of the $k$ points and $l_j(0)$ coefficients are combined, then the secret can be computed as:   
\begin{equation} \label{eq:find_m}
    M = \sum_{j=0}^{k-1} l_j(0) \cdot y_j,
\end{equation}
This method makes it easier to compute M without having to recalculate each $l_j(x)$.

\subsection{Anonymous Communication using Onion Routing} \label{ssec:onion_routing}

Onion routing is widely used in the domain of computer networks when routing has to be done while keeping the sender anonymous. Each message is encrypted several times (layers of encryption) analogous to layers of an onion. Each intermediate router from source to destination (called onion routers) ``peels" a single layer of encryption revealing the next hop. The final layer is decrypted and message is read at the destination. The identity of the sender is preserved since each intermediate router only knows the preceding and the following routers. The overhead of onion routing comes from the fact that the sender has to do several rounds of encryption before sending the packet to the network and each intermediate router has to do a decryption before forwarding it to the next hop. While this can be done in computer networks, adopting this in resource-constrained NoCs leads to unacceptable performance overhead as illustrated in Section~\ref{sec:motivation}.

\subsection{Related Work} \label{ssec:related_work}

The current state-of-the-art in NoC security revolves around protecting information traveling in the network against side channel~\cite{reinbrecht2016side}, physical~\cite{raparti2019lightweight} and software attacks~\cite{ancajas2014fort}. Other attacks such as denial-of-service~\cite{charles2020real,boraten2016secure} and buffer overflow~\cite{lukovic2010enhancing,fiorin2008security} have also been explored. However, developing efficient and flexible solutions at lower costs and minimal impact on performance as well as how to certify these solutions remain as challenges to the industry. It is not feasible to adopt the security mechanisms used in the computer networks domain in NoC based SoCs due to the resource-constrained nature of embedded devices~\cite{ogras2010analytical,charles2018exploration}. Security has to be considered in the context of other non-functional requirements such as performance, power and area. The obvious extension is to optimize these security mechanisms to fit the performance and power budgets of embedded systems. This thought process has led to prior efforts on securing NoC-based SoC~\cite{sajeesh2011authenticated,sepulveda2017towards,ancajas2014fort}, which tried to eliminate complex encryption schemes such as AES and replace them with lightweight encryption schemes.  Intel's TinyCrypt, a cryptographic library with a small footprint, is built for resource constrained devices~\cite{tinycrypt}. It provides basic functionality to build a secure system with minor overhead. It provides SHA-256 hash functions, message authentication, a psuedo-random number generator which can run using minimal memory, encryption, and the ability to create nonces\footnote{A {\bf nonce} is a random number that is used only once during the lifetime of a cryptographic operation.} and challenges. Apart form Intel TinyCrypt, several researches have proposed other lightweight encryption solutions in the Internet-of-Things (IoT) domain ~\cite{naru2017recent,babar2011proposed}. However, all of these solutions follow the traditional encryption method which takes a key and a plaintext as inputs to produce the ciphertext. The complex cryptographic operations required for such methods incur considerable overhead. In contrast, we propose a method where each router along the routing path contributes a portion of the message such that the message changes at each router and only the destination receives the entire message. This can be implemented using very few addition and multiplication operations leading to a lightweight solution for secure communication.

Existing work on anonymous routing (e.g., onion routing, mix-nets, dining cryptographers, etc.) considers mobile ad-hoc networks (MANETS)~\cite{kong2003anodr,qin2008olar,liu2014aasr} as well as computer and vehicular networks~\cite{yuan2014anonymous}. The idea behind the widely used \textit{onion routing} is explained in Section~\ref{ssec:onion_routing}. The main challenge in using these anonymous routing protocols in resource-constrained SoCs is that the protocol uses decryption (``peeling the onion") at each hop leading to unacceptable performance overhead. Optimized anonymous routing protocols in MANETS (e.g., ~\cite{qin2008olar}) use an on-demand lightweight anonymous routing protocol that eliminates per-hop decryption. However, the MANETS environment is fundamentally different from an NoC. Their work cannot address the unique communication requirements of an NoC as well as not designed for task-migration and context switching. {\it To the best of our knowledge, our work is the first attempt in developing an anonymous routing protocol for NoC based SoCs}.

\section{Motivation} \label{sec:motivation}

Security and performance is always a trade-off in resource-constrained systems. While computer networks with potentially unlimited resources can accommodate very strong security techniques such as AES encryption and onion routing, utilizing them in resource-constrained NoCs can lead to unacceptable overhead. To evaluate this impact, we ran FFT, RADIX (RDX), FMM and LU benchmarks from the SPLASH-2 benchmark suite \cite{woo1995splash} on an $8 \times 8$ Mesh NoC-based SoC with 64 IPs using the gem5 simulator \cite{binkert2011gem5} considering three scenarios:

\begin{itemize}
    \item {\it No-Security:} NoC does not implement encryption or anonymous routing.
    \item {\it Enc-only:} NoC secures data by encrypting before sending into the network. However, it does not support anonymous routing.
    \item {\it Enc-and-AR:} Data encryption as well as anonymous routing achieved by onion routing.
\end{itemize}

\begin{figure}[h]
\centering
\begin{subfigure}{0.8\columnwidth}
  \centering
  \includegraphics[width=1\linewidth]{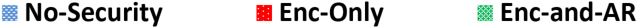}
\end{subfigure}\\%
\vspace{2mm}
\begin{subfigure}{0.8\columnwidth}
  \centering
  \includegraphics[width=1\linewidth]{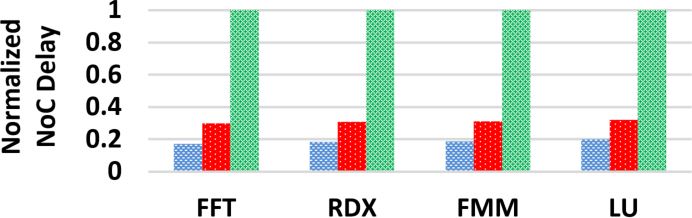}
  \caption{NoC delay}
  \label{fig:motiv-noc-delay}
\end{subfigure}%
\hspace{1mm}
\begin{subfigure}{0.8\columnwidth}
  \centering
  \includegraphics[width=1\linewidth]{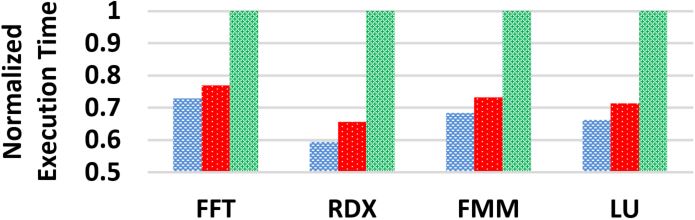}
  \caption{Execution time}
  \label{fig:motiv-exec-time}
\end{subfigure}
\caption{NoC delay and execution time comparison across different levels of security.}
\vspace{-4mm}
\label{fig:motiv}
\end{figure}

We assumed a 12-cycle delay for encryption/decryption when simulating Enc-only and Enc-and-AR according to the evaluations in~\cite{sajeesh2011authenticated}. More details about the experimental setup is given in Section~\ref{ssec:ex_setup}. Results are shown in Figure~\ref{fig:motiv}. The values are normalized to the scenario that consumes the most time. Enc-only shows 42\% (40\% on average) increase in NoC delay (total NoC traversal delay for all packets) and 9\% (7\% on average) increase in execution time compared to the No-Security implementation. Enc-and-AR gives worse results with 83\% (81\% on average) increase in NoC delay leading to a 41\% (33\% on average) increase in execution time when compared with No-Security. In other words, Enc-and-AR leads to approximately 1.5X performance degradation.  
When security is considered, No-Security leaves the data totally vulnerable to attackers, Enc-only secures the data by encryption and Enc-and-AR provides an additional layer of security with anonymous routing. The overhead of Enc-only is caused by the complex mathematical operations, and the number of cycles required to encrypt each packet. Onion routing used in Enc-and-AR aggravates this by requiring several rounds of encryption before injecting the packet into the network as well as decryption at each hop (router). Added security has less impact on execution time compared to NoC delay since execution time also includes the time for instruction execution and memory operations in addition to NoC delay. In many embedded systems, it would be unacceptable to have security at the cost of 1.5X performance degradation. It would be ideal if the security provided by Enc-and-AR can be achieved while maintaining performance comparable to No-security. Our approach tries to achieve this goal by introducing a lightweight encryption and anonymous routing protocol as described in the next section.

\section{Lightweight Encryption and Anonymous Routing Protocol} \label{sec:learn}

This section describes our proposed approach  - \textit{Lightweight Encryption and Anonymous Routing protocol for NoCs} (LEARN). By utilizing secret sharing based on \textit{polynomial interpolation}~\cite{shamir1979share}, LEARN negates the need for complex cryptographic operations to encrypt messages.  
A forwarding node would only have to compute the low overhead addition and multiplication operations to hide the contents of the message. 
As the message passes through the forwarding path, its appearance is changed at each node, which makes the message's content and route safe from eavesdropping attackers as well as internal ones. The following sections describe our approach in detail. First, we provide an overview of our framework in Section~\ref{ssec:overview}. Next, Section~\ref{ssec:route-discovery} and Section~\ref{ssec:data-transfer} describe the two major components of our proposed routing protocol (route discovery and data transfer). Finally, Section~\ref{ssec:parameter-mgt} outlines how to efficiently manage relevant parameters during anonymous routing.

\begin{figure*}[t]
    \centering
    \includegraphics[width=0.9\textwidth]{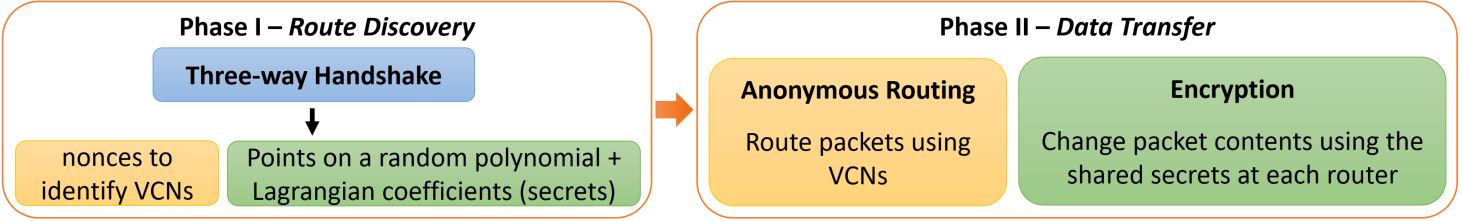}
    \caption{Overview of our proposed framework (LEARN)}
    \vspace{-2mm}
    \label{fig:summary} 
\end{figure*}

\subsection{Overview} \label{ssec:overview}

LEARN has two main phases as shown in Figure~\ref{fig:summary}. 
When an IP wants to communicate with another IP, it first completes the ``Route Discovery" phase. The route discovery phase sends a packet and discovers the route, distributes the parameters among participants. Then the ``Data Transfer" phase transfers the message securely and anonymously.
The route discovery phase includes a three-way handshake between the sender and the destination nodes. The handshake uses 3 out of the 4 main types of packets sent over the network with the fourth type being used in the second phase. The 4 main packet types are:

\begin{enumerate}
    \item $RI$ (Route Initiate) - flooded packet from sender $S$ to destination $D$ to initialize the conversation.
    \item $RA$ (Route Accept) - packet sent from $D$ to accept new connection with $S$.
    \item $RC$ (Route Confirmation) - sent from $S$ to distribute configuration parameters with intermediate nodes.
    \item $DT$ (Data) - the data packet from $S$ to $D$ that is routed anonymously through the NoC.
\end{enumerate}

Algorithm 1 outlines the major steps of LEARN. During the three-way handshake, a route between $S$ and $D$ is discovered. Each router along the routing path is assigned with few parameters that are used when transferring data - (i) random nonces to represent preceding and following routers (line 3), and (ii) a point in a random polynomial together with its Lagrangian coefficient (line 4). This marks the end of the first phase which enables the second phase - ``Data Transfer". The second phase uses the parameters assigned to each router to forward the original message through the route anonymously while hiding its contents. Anonymous routing is achieved by using the random nonces which act as \textit{virtual circuit numbers} (VCN). When transferring data packets, the intermediate routers will only see the VCNs corresponding to the preceding router and the following router which reveals no information about the source or the destination (line 8). Encryption is achieved using the points in the random polynomial and their corresponding Lagrangian coefficients. Each router along the path changes the contents of the message in such a way that only the final destination will be able to retrieve the entire message (line 7). \vspace{-2mm}

{\underline{\textbf{Algorithm 1} - Major steps of LEARN}} \\ \vspace{-4mm}
\begin{algorithmic}[1] \label{alg:overview} 
\State \textit{Phase I - Route Discovery}
\ForAll{$r \in$ routers}
\State $r \leftarrow \upsilon_i, \upsilon_j$ \Comment{nonces to identify VCNs}
\State $r \leftarrow (x_k, y_k, b_k)$ \Comment{a point in a random polynomial}
\EndFor \vspace{1mm}
\State \textit{Phase II - Data Transfer}
\While{$r \neq destination$}
\State $m \leftarrow \calF(m, (x_k, y_k, b_k))$ \Comment{modify message}
\State $r \leftarrow getNextHop(\upsilon_i, \upsilon_j)$ \Comment{get next hop}
\EndWhile
\end{algorithmic}

LEARN improves performance by replacing complex cryptographic operations with addition/multiplication operations that consume significantly less time during the data transfer phase. The overhead occurs during the first phase (route discovery) that requires cryptographic operations. However, this is performed only a constant number of times (once per communication session). Since the route discovery phase happens only once in the beginning of a communication session, the cost for route discovery gets amortized over time. This leads to significant performance improvement.

Note that the route discovered at the route discovery stage will remain the same for the lifetime of the task. In case of context switching and/or task migration, the first phase will be repeated before transferring data. Each IP in the SoC that uses the NoC to communicate with other IPs follows the same procedure. The next two sections describe these two phases in detail. A list of notations used to illustrate the idea is listed in Table~\ref{tab:notations}. The superscript ``$i$" is used to indicate that the parameter is changed for each packet of a given packet type.

\begin{table}[H]
\caption{Notations used to illustrate LEARN} \label{tab:notations}
\begin{tabularx}{\columnwidth}{|p{1cm}|X|}
\hline
$OPK_S^{(i)}$ & one-time public key (OPK) used by the  source to uniquely identify an $RA$ packet \\ \hline
$OSK_S^{(i)}$  & \begin{tabular}[c]{@{}l@{}}private key corresponding to $OPK_S^{(i)}$\end{tabular}                          \\ \hline
$\rho$ & random number generated by the source                                                                     \\ \hline
$PK_D$ & the global public key of the destination                                                                    \\ \hline
$SK_D$ & \begin{tabular}[c]{@{}l@{}}the private key corresponding to $PK_D$\end{tabular}                         \\ \hline
$TPK_A^{(i)}$  & temporary public key of node $A$                                                                \\ \hline
$TSK_A^{(i)}$  & \begin{tabular}[c]{@{}l@{}}the private key corresponding to $TPK_A^{(i)}$ \end{tabular}                            \\ \hline
$K_{S-A}$  & symmetric key shared between $S$ and $A$                                                           \\ \hline
$\upsilon_A$ & randomly generated nonce by node $A$                                                          \\ \hline
$b_i$ & \begin{tabular}[c]{@{}l@{}}Lagrangian coefficient of a given point $(x_i,y_i)$\end{tabular} \\ \hline
$E_K(M)$ & a message $M$ encrypted using the key $K$ \\
\hline
\end{tabularx}
\end{table}


\subsection{Route Discovery} \label{ssec:route-discovery}

\begin{figure*}[h]
    \centering
    \includegraphics[width=1\textwidth]{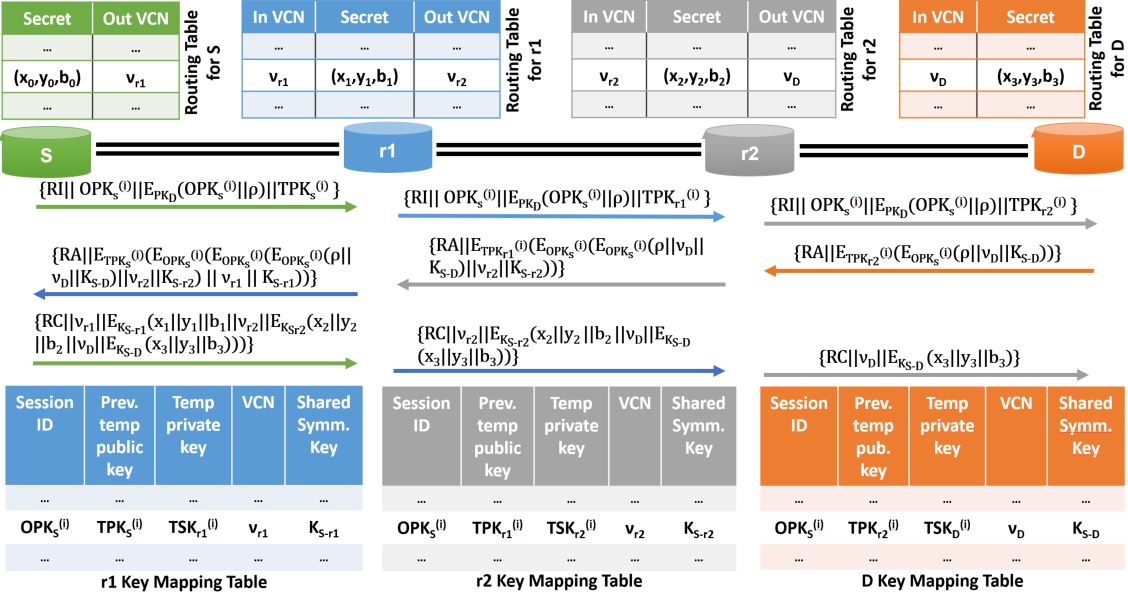}
    \caption{Steps of the three-way handshake and the status of parameters at the end of the process.}
    \label{fig:learn-handhsake}
\end{figure*}

The route discovery phase performs a three-way handshake between the sender $S$ and destination $D$. This includes broadcasting the first packet - RI from $S$ with the destination $D$, getting a response (RA) from $D$ acknowledging the reception of RI, and finally, sending RC with the parameters required to implement polynomial interpolation based secret sharing. Figure~\ref{fig:learn-handhsake} shows an illustrative example of parameters (using only four nodes) shared and stored during the handshake.

The initial route initiate packet ($RI$) takes the form: 
\begin{equation*}
    \{RI \concat OPK_S^{(i)} \concat E_{PK_D} (OPK_S^{(i)} \concat \rho ) \concat TPK_S^{(i)} \}
\end{equation*}

The first part of the message indicates the type of packet being sent, $RI$ in this case. $OPK_S^{(i)}$ refers to the one-time public key associated with the sender node. This public key together with its corresponding private key $OSK_S^{(i)}$ change with each new conversation or $RI$. This change allows for a particular conversation to be uniquely identified by these keys, which are saved in its \textit{route request table}.  $\rho$ is a randomly generated number by the sender that is concatenated with the $OPK_S^{(i)}$ and then encrypted with the destination node’s public key $PK_D$ as a global trapdoor~\cite{katz1996handbook}. Since $PK_D$ is used to encrypt, only the destination is able to open the trapdoor using $SK_D$. Then the $TPK_S^{(i)}$ is attached to show the temporary key of the forwarding node, which is initially the sender. The temporary keys are also implemented as one-time trapdoors to ensure security.

The next node, $r1$, to receive the $RI$ messages goes through a few basic steps.  Firstly, it checks for the $OPK_S^{(i)}$ in its \textit{key mapping table}, which would indicate a duplicated message. Any duplicates are discarded at this step. Next, $r1$ will attempt to decrypt the message and retrieve $\rho$. Success would indicate that $r1$ was the intended recipient $D$. If not, $r1$ replaces $TPK_S^{(i)}$ with its own temporary public key $TPK_{r1}^{(i)}$ and broadcasts: 
\begin{equation*}
    \{RI \concat OPK_S^{(i)} \concat E_{PK_D} (OPK_S^{(i)} \concat \rho ) \concat TPK_{r1}^{(i)} \}
\end{equation*}
$r1$ also logs $OPK_S^{(i)}$ and $TPK_S^{(i)}$ from the received message and $TSK_{r1}^{(i)}$ corresponding to $TPK_{r1}^{(i)}$ in its key mapping table. This information is used later when an $RA$ message is received from $D$.

$D$ will eventually receive the $RI$ message and will decrypt using $SK_D$. This will allow $D$ to retrieve $OPK_S^{(i)}$ and $\rho$ from $E_{PK_D} (OPK_S^{(i)} \concat \rho )$. Then to verify that the $RI$ has not been tampered with, $D$ will compare the plaintext $OPK_S^{(i)}$ and the now decrypted $OPK_S^{(i)}$.  If they are different, the $RI$ is simply discarded. Otherwise, $D$ sends a $RA$ (route accept) message: 
\begin{equation} \label{eq:ra_init}
    \{RA \concat E_{TPK_{r2}^{(i)}} ( E_{OPK_S^{(i)}} (\rho \concat \upsilon_D \concat K_{S-D} )) \}
\end{equation}
$RA$, like $RI$ in the previous message, is there to indicate message type.  $D$ generates a random nonce, $\upsilon_D$, to serve as a VCN and a randomly selected key $K_{S-D}$ to act as a symmetric key between $S$ and $D$. $D$ stores $\upsilon_D$ and $K_{S-D}$ in its key mapping table. It also makes an entry in its routing table indexed by $\upsilon_D$, the VCN. The concatenation of $\rho$, $\upsilon_D$, and $K_{S-D}$ is then encrypted with the $OPK_S^{(i)}$, so that only $S$ can access that information.  Then the message is encrypted again by $TPK_{r2}^{(i)}$, $r2$’s temporary public key, with $r2$ being the node that delivered $RI$ to $D$. 

Once $r2$ receives the $RA$, it decrypts it using its temporary private key, $TSK_{r2}^{(i)}$, and follows the same steps as $D$. It generates its own nonce, $\upsilon_{r2}$, and shared symmetric key, $K_{S-r2}$, to be shared with $S$. Both the nonce and symmetric key are then concatenated to the $RA$ message and encrypted by $S$’s public key, $OPK_S^{(i)}$, so that only $S$ can retrieve that data. This adds another layer of encrypted content to the message for $S$ to decrypt using $OSK_S^{(i)}$. Similar to $D$, $r2$ also stores $\upsilon_{r2}$ and $K_{S-r2}$ in its key mapping table and routing table. It then finds the temporary public key for the previous node in the path from its key mapping table - $TPK_{r1}^{(i)}$ and encrypts the message. The message sent out by $r2$ looks like:
\begin{multline} \label{eq:ra_by_r2}
    \{RA \concat E_{TPK_{r1}^{(i)}} (E_{OPK_S^{(i)}}( E_{OPK_S^{(i)}} (\rho \concat \upsilon_D \concat K_{S-D} ) \\ \concat \upsilon_{r2} \concat K_{S-r2})) \}
\end{multline}

This process is repeated at each node along the path until the $RA$ packet makes it way back to $S$. The entire message at that point is encrypted with $TPK_{S}^{(i)}$, which is stripped away using $TSK_{S}^{(i)}$. Then $S$ can ``peel" each layer of the encrypted message by $OSK_S^{(i)}$ to retrieved all the VCNs, shared symmetric keys, and also, $\rho$.  $\rho$ is used to authenticate that the entire message came from the correct destination and was not changed during the journey.

\begin{figure}[h]
    \centering
    \includegraphics[width=0.8\columnwidth]{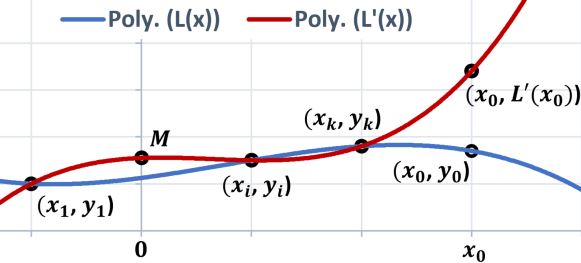}
    \caption{Lagrangian polynomials $L(x)$ and $L'(x)$ together with the selected points.}
    \vspace{-2mm}
    \label{fig:polynomials} 
\end{figure}

Once $S$ completes authentication of the received RA packet, it randomly generates $k+1$ points $(x_0,y_0), (x_1, y_1), ..., (x_k,y_k)$ on a $k$ degree polynomial $L(x)$ as shown in Figure~\ref{fig:polynomials}. $k+1$ is the number of nodes in the path from $S$ to $D$. $S$ then uses these points to calculate the Lagrangian coefficients, $b_0, b_1, ... , b_k$, using: 
\begin{equation} \label{eq:lag_coef_s}
    b_j = \prod_{i=0, i \neq j}^{k} \frac{x_i}{x_i - x_j}
\end{equation}

Using the generated data, $S$ constructs a route confirmation (RC) packet:
\begin{multline} \label{eq:rc_msg}
    \{RC \concat \upsilon_{r1} \concat E_{K_{S-r1}}(x_1 \concat y_1 \concat b_1 \concat \upsilon_{r2} \concat E_{K_{S-r2}}(x_2 \concat y_2 \\ \concat b_2 \concat \upsilon_{D} \concat E_{K_{S-D}} (x_3 \concat y_3 \concat b_3)) ) \}
\end{multline}
Similar to the case in RA and RI, RC in the packet refers to the packet type. The rest of the message is layered much like the previous $RA$ packet. Each layer contains the $\upsilon_*$  for each node  concatenated with secret information that is encrypted with the shared key $K_{S-*}$, where * corresponds to $r1, r2$ or $D$ in our example (Figure~\ref{fig:learn-handhsake}). The ($\upsilon_*$, $K_{S-*}$) pair was generated by each node during the RA packet transfer phase and the values were stored in the key mapping tables as well as entries indexed by the VCNs created in the routing table. Therefore, each node can decrypt one layer, store incoming and outgoing VCNs together with the secret, and pass it on to the next node to do the same.
For example, $r1$ receiving the packet can observe that the incoming VCN is $\upsilon_{r1}$. It then decrypts the first layer using the symmetric key $K_{S-r1}$, that is already stored in the key mapping table, and recovers the secret $(x_1, y_1, b_1)$ as well as the outgoing VCN $\upsilon_{r2}$. It then updates the entry indexed by $\upsilon_{r1}$ in its routing table with the secret tuple and the outgoing VCN. Similarly, each router from $S$ to $D$ can build its routing table.

\subsection{Data Transfer} \label{ssec:data-transfer}

The path set up can now be used to transfer messages from $S$ to $D$ anonymously. For each conversation, $k+1$ points were generated on a random curve $L(x)$ chosen by $S$. During the last step of the route discovery phase (RC packet), $S$ kept $(x_0, y_0, b_0)$ for itself and distributed each node on the discovered path a different point, $(x_i, y_i)$ (where $1 \leq i \leq k$), with the corresponding Lagrangian coefficient $b_i$. If $S$ wants to send the message $M$ to $D$, $S$ has to generate a new $k$ degree polynomial $L'(x)$ which is defined by the $k$ points distributed to nodes except for $(x_0,y_0)$, i.e., points $(x_i, y_i)$ where $(1 \leq i \leq k)$ and a new point $(0,M)$. This makes $L'(0) = M$ with $M$ as the secret message, according to the explanation in Section~\ref{ssec:intro_ss}. $S$ then changes its own point $(x_0, y_0)$ to $(x_0, y'_0$) where $y'_0 = L'(x_0)$, making sure the point retained by $S$ is also on the curve L'(x) as shown in Figure~\ref{fig:polynomials}. It is important to note that every coefficient $b_i$, and every point distributed to nodes along the route remain unchanged. For this scenario, considering Equation~\ref{eq:find_m}, we can derive:
\begin{equation} \label{eq:find_new_m}
    M = y'_0b_0 + \sum_{i=1}^{k} b_i \cdot y_i
\end{equation}
To transfer a secret message, $M$, from $S$ to $D$ anonymously, $S$ constructs data transfer ($DT$) packet with the form:
\begin{equation}
    \{ DT \concat \upsilon_{r1} \concat y'_0b_0 \}
\end{equation}
$DT$, like every other packet, has an indicator of packet type at the front of the packet - $DT$. $\upsilon_{r1}$ is the VCN of the next node. $y'_0 b_0$ is the portion of the message $M$ that is constructed by $S$.  Once $r1$ receives the $DT$ packet, it adds its own portion of the message, $y_1 b_1$, to $y'_0 b_0$. It also uses its routing table to find the VCN of the next node and replaces the incoming VCN by the outgoing VCN in the $DT$ packet. Therefore, the message received by $r2$ has the form:
\begin{equation}
    \{ DT \concat \upsilon_{r2} \concat y'_0b_0 + y_1b_1 \}
\end{equation}
Next, $r2$ repeats the same process and forwards the packet:
\begin{equation}
    \{ DT \concat \upsilon_{D} \concat y'_0b_0 + y_1b_1 + y_2b_2 \}
\end{equation}
to $D$. Eventually, $D$ will be able to retrieve the secret message, $M = y'_0b_0 + y_1b_1 + y_2b_2 + y_3b_3$ by adding the last portion $y_3b_3$ constructed using the part of the secret $D$ shared. Using this method, neither an intermediate node nor an eavesdropper in the middle will be able to see the full message since the message $M$ is incomplete at every intermediate node and is fully constructed only at the destination $D$.

\subsection{Parameter Management} \label{ssec:parameter-mgt}

To ensure the efficient implementation of LEARN, an important aspect needs to be addressed - the generation and management of keys and nonces. However, this is beyond the scope of this paper and many previous studies have addressed this problem in several ways. One such example is the work done by Lebiednik et al.~\cite{lebiednik2018architecting}. In their work, a separate IP called the \textit{key distribution center} (KDC) handles the distribution of keys. Each node in the network negotiates a new key with the KDC using a pre-shared portion of memory that is known by only the KDC and the corresponding node. The node then communicates with the KDC using this unique key whenever it wants to obtain a new key. The KDC can then allocate keys depending on whether it is symmetric/asymmetric encryption, and inform other nodes as required. The key request can delay the communication. But once keys are established, it can be used for many times depending on the length of the encrypted packet before refreshing to prevent linear distinguishing attacks. In our approach, the keys are only used during the route discovery phase, and the discovered route will remain the same for the lifetime of the task unless context switching or task migration happens. Therefore, key refreshing will rarely happen and the cost for the initial key agreement as well as the route discovery phase will be amortized.


\color{black}


\section{Experimental Results} \label{sec:results}

This section presents results to evaluate the efficiency of our approach (LEARN). We first describe the experimental setup. Next, we compare the performance of LEARN with traditional encryption and anonymous routing protocols introduced in Section~\ref{sec:motivation}. Finally, we discuss the area overhead and security aspects of LEARN.

\subsection{Experimental Setup} \label{ssec:ex_setup}

Extending the results presented in Figure~\ref{fig:motiv}, LEARN was tested on an $8 \times 8$ Mesh NoC-based SoC with 64 IPs using the gem5 cycle-accurate full-system simulator~\cite{binkert2011gem5}. The NoC was built using the ``GARNET2.0" model that is integrated with gem5~\cite{agarwal2009garnet,charles2018exploration}. 
The route discovery phase of our approach relies on the $RI, RA,$ and $RC$ packets traversing along the same path to distribute the keys and nonces. Therefore, the topology requires bidirectional links connecting the routers. While we experimented on a Mesh NoC, there are many other NoC topologies that can adopt LEARN where all links are bidirectional as evidenced by academic research~\cite{agarwal2009garnet} as well as commercial SoCs~\cite{sodani2016knights}.

Each encryption/decryption is modelled with a 12-cycle delay~\cite{sajeesh2011authenticated}. Computations related to generating the random polynomial and deciding the $k$ points is assumed to consume 200 cycles.
To accurately capture congestion, the NoC was modeled with 3-stage (buffer write, route compute + virtual channel allocation + switch allocation, and link traversal) pipelined routers with wormhole switching and 4 virtual channel buffers at each input port. Each link was assumed to consume one cycle to transmit packets between neighboring routers.
The delays were chosen to be consistent with the delays of components in the gem5 simulator. 

We used the default gem5 and Garnet2.0 configurations for packet sizes, virtual channels and flow control. In addition to the four main types of packets described in Section~\ref{ssec:overview}, the $DT$ packets can be further divided into two categories as control and data packets. For example, in case of a cache miss, a memory request packet (control packet) is injected into the NoC and the memory response packet (data packet) consists of the data block from the memory. The address portion of a control $DT$ packet consists of 64 bits. In the data $DT$ packet, in addition to the 64-bit address, 512 bits are reserved for the data block. A credit-based, virtual channel flow control was used in the architecture. Each data VC and control VC was allocated buffer depths of 4 and 1, respectively.

\begin{figure}[h]
    \centering
    \includegraphics[width=0.9\columnwidth]{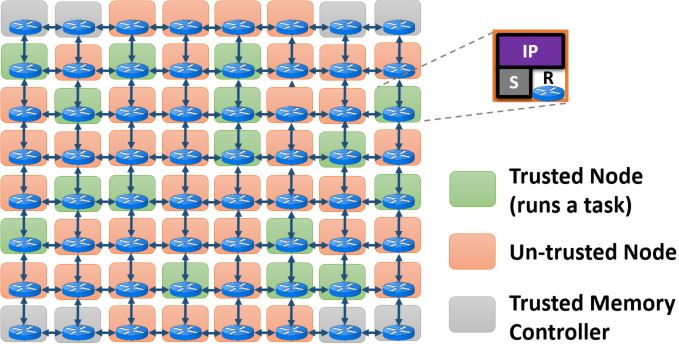}
    \caption{$8 \times 8$ Mesh NoC architecture used to generate results including trusted nodes running the tasks and communicating with memory controllers while untrusted nodes can potentially have malicious IPs.}
    \vspace{-2mm}
    \label{fig:ex_setup} 
\end{figure}

LEARN was tested using 6 real benchmarks (FFT, RADIX, FMM, LU, OCEAN, CHOLESKY) from the SPLASH-2 benchmark suite and 6 synthetic traffic patterns: \textit{uniform random (URD), tornado (TRD), bit complement (BCT), bit reverse (BRS), bit rotation (BRT), transpose (TPS)}.
\color{black}
Out of the 64 cores, 
16 IPs were chosen at random and each one of them instantiated an instance of the task. The packets injected into the NoC when running the real benchmarks were the memory requests/responses. We used 8 memory controllers that provide the interface to off-chip memory which were placed on the boundary of the SoC. This memory controller placement adheres to commercial SoC architectures such as Intel's Knights Landing (KNL)~\cite{sodani2016knights}. An example to illustrate the IP placement is shown in Figure~\ref{fig:ex_setup}.

When running real benchmarks, the packets get injected to the NoC when there are private cache misses and the frequency of that happening depends on the characteristics of the benchmark. When running synthetic traffic patterns, packets were injected into the NoC at the rate of 0.01 packets/node/cycle.
For synthetic traffic patterns, the destinations of injected packets were selected based on the traffic pattern. For example, \textit{uniform random} selected the destination from the remaining IPs with equal probability whereas \textit{bit complement}, complemented the bits of the source address to get the destination address, etc.. The choices made in the experiments were motivated by the architecture/threat model and the behavior of the gem5 simulator. However, LEARN can be used with any other NoC topology and task/memory controller placement.

\subsection{Performance Evaluation} \label{ssec:performance-evaluation}

\begin{figure}[h]
\centering
\begin{subfigure}{0.85\columnwidth}
  \centering
  \includegraphics[width=1\linewidth]{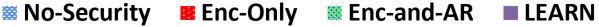}
\end{subfigure}\\%
\vspace{1mm}
\begin{subfigure}{1\columnwidth}
  \centering
  \includegraphics[width=1\linewidth]{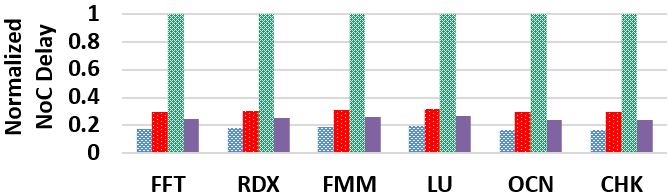}
  \caption{NoC delay}
  \label{fig:results-noc-delay}
\end{subfigure}%
\hspace{1mm}
\begin{subfigure}{1\columnwidth}
  \centering
  \includegraphics[width=1\linewidth]{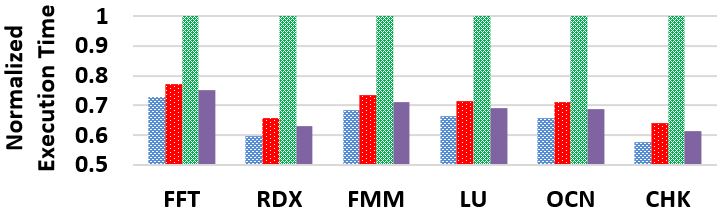}
  \caption{Execution time}
  \label{fig:results-exec-time}
\end{subfigure}
\caption{NoC delay and execution time comparison across different security levels using real benchmarks.}
\label{fig:performance-real}
\end{figure}

Figure~\ref{fig:performance-real} shows performance improvement LEARN can gain when running real benchmarks. We compare the results from LEARN against the three scenarios considered in Figure~\ref{fig:motiv}. 
Compared to the No-Security scenario, LEARN consumes 30\% more time (28\% on average) for NoC traversals (NoC delay) and that results in only 5\% (4\% on average) increase in total execution time. Compared to Enc-and-AR which also implements encryption and anonymous routing, LEARN improves NoC delay by 76\% (74\% on average) and total execution time by 37\% (30\% on average). We can observe from the results that the performance of LEARN is even better than Enc-Only, which provides encryption without anonymous routing. Overall, LEARN can provide encryption and anonymous routing consuming only 4\% performance overhead compared to the NoC that does not implement any security features.

\begin{figure}[h]
    \centering
    \includegraphics[width=1\linewidth]{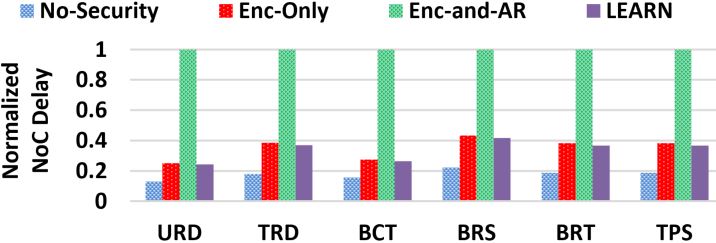}
    \caption{NoC delay comparison across different levels of security when running synthetic traffic patterns.}
    \label{fig:performance-synth}
\end{figure}

The same experiments were carried out using synthetic traffic traces, and results are shown in Figure~\ref{fig:performance-synth}. Since synthetic traffic patterns only simulate NoC traffic and do not include instruction execution and memory operations, only NoC delay is shown in the figure. Compared to Enc-and-AR, LEARN improves performance by 76\% (72\% on average).

The performance improvement of LEARN comes from the fact that once the path has been set up for the communication between any two IPs, the overhead caused to securely communicate between the two IPs (data transfer phase) while preserving route anonymity is much less. The notable overhead occurs at the route discovery phase due to complex cryptographic operations. The intermediate nodes encrypt/decrypt packets to exchange parameters securely. Yet, these complex cryptographic operations are performed only a constant number of times. Majority of the work is done at the source which selects points to be distributed among intermediate nodes after constructing a curve, calculates the Lagrangian coefficients of the selected points, and performs several rounds of encryption/decryption during the three-way handshake. Once the routing path is setup, packets can be forwarded from one router to the other by a simple table look-up. No per-hop encryption and decryption is required to preserve anonymity. The security of a message is ensured by changing the original message at each node using a few addition and multiplication operations which incur significantly fewer extra delays. Since the route discovery phase happens only once during the lifetime of a task unless context switching and/or task migration happens, and there is only a limited number of communications going on between IPs in an SoC, the cost during the route discovery phase gets amortized over time. 
When running real benchmarks, we observed a packet ratio of 1:1:1:6325 on average for $RI:RA:RC:DT$, respectively. For synthetic traffic patterns, the same ratio was observed to be 1:1:1:1964.
This leads to a significant performance improvement compared to the traditional methods of encryption and anonymous routing.

\subsection{Area Overhead of the Key Mapping Table} \label{ssec:key-mapping-overhead}

The key mapping table is an extra table compared to No-Security approach used to implement our anonymous routing protocol.
The key mapping table adds a row for each session. Therefore, the size of the key mapping table is linearly proportional to the number of sessions. If at design time, it is decided to have a fixed size for the key mapping table, it is possible for the key mapping table at a router to be full after adding sessions, and in that case, new sessions cannot be added through that router. Therefore, the size has to be decided according to the communication requirements.
    
The maximum number of communication pairs in an $8 \times 8$ Mesh is $\binom{64}{2} \times 2 = 4032$ (assuming two-way communication between any pair out of the 64 nodes). Depending on the address mapping, only some node pairs (out of all the possible node pairs) communicate. Our simulations consisted of 256 unique node pairs. 
In the worst case, if we assume each communication session has one common router, the key mapping table should be $256 \times row\_size$ big. If each entry in the key mapping table is 128 bits, the total size becomes 20kB. However, in reality, not all communication sessions overlap. It is also important to note that except for the Session ID in the key mapping table, the other entries can be overwritten once route discovery phase is complete.
Therefore, it is possible to allocate a fixed size key mapping table during design time and yet keep the area overhead low.

\subsection{Security Analysis} \label{ssec:security}

In this section, we discuss the security and privacy of messages transferred on the NoC using LEARN.

{\it Security of messages: }
The security of messages is preserved by the $(k,n)$ threshold property of Lagrangian polynomials discussed in Section~\ref{ssec:intro_ss}. Therefore, unless an intermediate node can gather all points distributed among the routers in the routing path together with their Lagrangian coefficients, the original message cannot be recovered. Our threat model states that the source and destination are trusted IPs, and also, only some of the IPs are untrusted. Therefore, all routers along the routing path will never be compromised at the same time. The threat comes from malicious IPs sitting on the routing path and eavesdropping to extract security critical information. LEARN ensures that intermediate nodes that can be malicious, cannot recover the original message during the data transfer phase by changing the message at each hop. The complete message can only be constructed at the destination. During route discovery phase, each packet is encrypted such that only the intended recipient can decrypt it. The key and nonce exchange is also secured according to the mechanism proposed in Section~\ref{ssec:parameter-mgt}. Therefore, LEARN ensures that no intermediate M3PIP can gather enough data to recover the plaintext from messages. 

{\it Anonymity of nodes in the network: }
LEARN preserves the anonymity of nodes in the network during all of its operational phases. When the source sends the initial RI packet to initiate the three-way handshake, it doesn't use the identity of the destination. Instead, the source uses the global public key of the destination ($PK_D$) and sends a broadcast message on the network. When the RI packet propagates through the network, each intermediate node saves a temporary public key of its predecessor. This temporary public key is then used to encrypt data when propagating the RA packet so that unicast messages can be sent to preceding nodes without using their identities. Random nonces and symmetric keys are assigned to each node during the RA packet propagation which in turn is used by the RC packet to distribute points and Lagrangian coefficients to each node. Data transfer is done by looking up the routing table that consists of the nonces representing incoming and outgoing VCNs. Therefore, the identities of the nodes are not revealed at any point during communication.

{\it Anonymity of routes taken by packets: }
In addition to preserving the anonymity of nodes, LEARN also ensures that the path taken by each packet is anonymous. Anonymity of the routing path is ensured by two main characteristics. (i) The message is changed at each hop. Therefore, even if there are two M3PIPs on the same routing path, information exchange among the two M3PIPs will not help in identifying whether the same message was passed through both of them. The same message appears as two completely different messages when passing through two different nodes. (ii) The routing table contains only the preceding and following nodes along the routing path. An M3PIP compromising a router will only reveal information about the next hop and the preceding hop. 
Therefore, the routing paths of all packets remain anonymous.


\section{Discussion} \label{sec:discussion}

In this section, we discuss possible alternatives to our design choices from both design overhead and security perspectives. Most importantly, we discuss security solutions to defend against attacks when an attacker is aware of our security mechanism.

\subsection{Feasibility of a Separate Service NoC} \label{ssec:service-noc}

Modern SoCs use multiple physical NoCs to carry different types of packets~\cite{wentzlaff2007chip,sodani2016knights}.
The KNL architecture used in Intel Xeon-Phi processor family uses four parallel NoCs~\cite{sodani2016knights}. The Tilera TILE64 architecture uses five Mesh NoCs, each used to transfer packets belonging to a certain packet type such as main memory, communication with I/O devices, and user-level scalar operand and stream communication between tiles~\cite{wentzlaff2007chip}.
The decision to implement separate physical NoCs is dependent on the performance versus area trade-off. If only one physical NoC is used to carry all types of packets, the packets must contain header fields such as $RI, RA, RC, DT$ to distinguish between different types. The buffer space is shared between different packet types. The SoC performance can deteriorate significantly due to these factors coupled with the increasing number of IPs in an SoC. On the other hand, contrary to intuition, due to the advancements in chip fabrication processes, additional wiring between nodes incur minimal overhead as long as the wires stay on-chip. 
Furthermore, when wiring bandwidth and on-chip buffer capacity is compared, the more expensive and scarce commodity is the on-chip buffer area. If different packet types are carried on NoC using virtual channels and buffer space is shared~\cite{diguet2007noc}, the increased buffer spaces and logic complexity to implement virtual channels becomes comparable to another physical NoC. A comprehensive analysis of having virtual channels versus several physical NoCs is given in~\cite{yoon2013virtual}.

It is possible to use two physical NoCs - one for data ($DT$) packet transfers and the other to carry packets related to the handshake ($RI, RA, RC$). However, in our setup, the potential performance improvement from a separate service NoC was not enough to justify the area and power overhead. We envision that our security mechanism to be a part of a suite of NoC security countermeasures that can address other threat models such as denial-of-service, buffer overflow, etc. The service NoC will be effective in such a scenario where more service type packets (e.g., DoS attack detection related packets~\cite{charles2019real}) are transferred through the NoC.

\subsection{Obfuscating the Added Secret} \label{ssec:hide-secret}

An attacker who is aware of our security mechanism can try to infer a communication path by observing the incoming and outgoing packets at a router.

Since each intermediate node adds a constant value ($y_ib_i$) to the received DT packet, the difference between incoming and outgoing DT packets at each node will be the same for a given virtual circuit.
For this attack to take place, two consecutive routers have to be infected by attackers and they have to collaborate. Alternatively, a Trojan in a router has to have the ability to observe both incoming and outgoing packets at the router. While these are strong security assumptions, it is important to address this loophole. In this Section, we propose a countermeasure against such an attack.
Even in the presence of such an attack, the secret message cannot be inferred since the complete message is only constructed at the destination and according to our threat model, we assume that the source and destination IPs are trustworthy.

This can be solved by changing the shared secret at each node for each message. However, generating and distributing secrets for each node per message can incur significant performance overhead. Therefore, we propose a solution based on each node updating its own secret.
According to Equation~\ref{eq:lag_coef_s}, to derive a new Lagrangian coefficient $b_i$, the $x$ coordinates should be changed. The source can easily do it for each message by changing both $x_0$ and $y_0$ when a new message needs to be sent. In other words, rather than changing the point ($x_0$,$y_0$) to ($x_0$,$y_0'$), it should be changed to ($x_0'$,$y_0'$). However, the new $x_0'$ now has to be sent to each intermediate node for them to be able to calculate the new secrets using:
\begin{equation}
    b'_j = b_j \cdot \frac{x'_0}{x'_0 - x_j} \cdot \frac{x_0 - x_j}{x_0}
\end{equation}
We want to avoid such communications for performance as well as security concerns. An alternative is to use a function $\mathcal{F}(x_0,\delta)$ that can derive the next $x$-coordinate starting from the initial $x_0$.
\begin{equation}
    x_0' = \mathcal{F}(x_0,\delta)
\end{equation}
where $\mathcal{F}(x_0,\delta)$ can be a simple incremental function such as $\mathcal{F}(x_0,\delta) = x+\delta$. $\delta$ can be a constant. To increase security, $\delta$ can be picked using a psuedo-random number generator (PRNG) seeded with the same value at each iteration. Using such a method will change the shared secrets at each iteration and that will remove correlation between incoming and outgoing packets at a node.

\subsection{Hiding the Number of Layers} \label{ssec:hide-onion}

Another potential vulnerability introduced by our approach is that attackers who are aware of our protocol, can infer how far they are from the source and destination based on the size of the RA and RC packets.
However, except for the corner case where the source/destination are at the edge of a certain topology, there can be more than one choice for potential source/destination candidates. In our experiments, we use the Mesh topology in which from the perspective of any node, there can be more than one node that is at distance $d$ away. However, the attacker can reduce the set of possible source/destination candidates for a given communication stream. 
Therefore, depending on the security requirements, this vulnerability can be addressed using the mechanism proposed in this section.

After receiving the RI Packet, when the $RA$ packet is initiated at $D$, $D$ generates $m$ $\langle$nonce, key$\rangle$ pairs ($ \langle \upsilon_D^1,K_{S-D}^1 \rangle , \langle \upsilon_D^2,K_{S-D}^2 \rangle,...,\langle \upsilon_D^m,K_{S-D}^m \rangle$) and adds $m$ layers to the packet. As a result, the $RA$ packet sent from $D$ to $r2$ takes the form:
\begin{multline*}
    \{RA \concat E_{TPK_{r2}^{(i)}} ( 
    E_{OPK_S^{(i)}}(... 
    E_{OPK_S^{(i)}}( 
    E_{OPK_S^{(i)}} (\rho \concat \upsilon_D^1 \\ \concat K_{S-D}^1 ) 
    \concat \upsilon_D^2 \concat K_{S-D}^2) ...
    \concat \upsilon_D^m \concat K_{S-D}^m)
    ) \}
\end{multline*}
$D$ stores the $\langle$nonce, key$\rangle$ pairs in its key mapping table. When $S$ receives the $RA$ packet, $S$ cannot distinguish whether the $m$ pairs were generated from multiple nodes or one node. Therefore, when the $RC$ packet is generated at $S$, instead of generating $k+1$ points (corresponding to the number of nodes in the path), the number of generated points depends on the number of $\langle$nonce, key$\rangle$ pairs received. During $RC$ packet transfer, each intermediate node along the routing path stores points (VCNs and secrets) corresponding to the nonces stored in the key mapping table. As a result, nodes can receive multiple secrets which can then be used during the data transfer phase. Depending on the required level of security, $m$ can vary and also, each intermediate node can add multiple layers to the $RA$ packet. 

\begin{figure}[h]
\centering
\begin{subfigure}{0.85\columnwidth}
  \centering
  \includegraphics[width=1\linewidth]{images/results-legend}
\end{subfigure}\\%
\vspace{1mm}
\begin{subfigure}{1\columnwidth}
  \centering
  \includegraphics[width=1\linewidth]{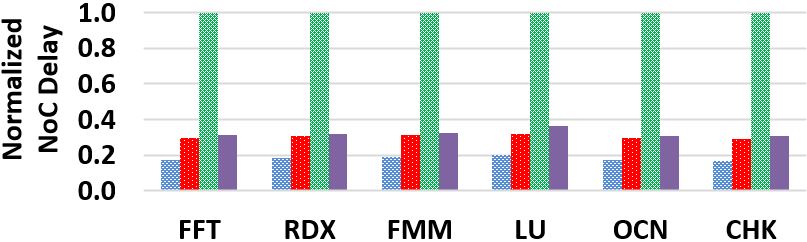}
  \caption{NoC delay}
  \label{fig:results-improve-noc-delay}
\end{subfigure}%
\hspace{1mm}
\begin{subfigure}{1\columnwidth}
  \centering
  \includegraphics[width=1\linewidth]{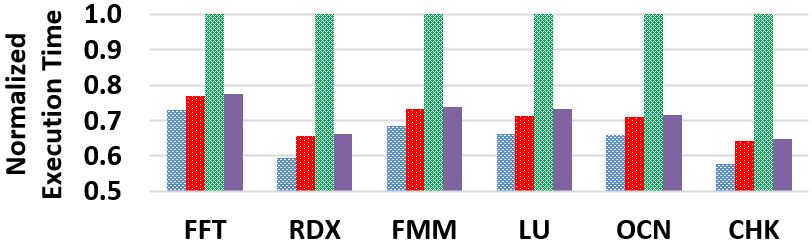}
  \caption{Execution time}
  \label{fig:results-improve-exec-time}
\end{subfigure}
\caption{NoC delay and execution time comparison across different security levels using real benchmarks considering the enhanced security features outlined in Section~\ref{ssec:hide-secret} and Section~\ref{ssec:hide-onion}.}
\label{fig:improve-performance-real}
\end{figure}

This method hides the correlation between the number of nodes and the length of the routing path, and therefore, eliminates the said vulnerability. However, this increases the performance penalty. Figure~\ref{fig:improve-performance-real} shows an extension of Figure~\ref{fig:performance-real} which considers the modification proposed in Section~\ref{ssec:hide-secret} and Section~\ref{ssec:hide-onion}.
LEARN improves NoC delay by 69\% (67\% on average) and total execution time by 34\% (27\% on average). Comparing with the results in Section~\ref{ssec:performance-evaluation}, the average total execution time improvement has been reduced by 3\% (from 30\% on average to 27\% on average) to accommodate the added security. 
Even then, LEARN enables significant performance improvement compared to traditional approaches.

\vspace{-2mm}
\section{Conclusions} \label{sec:conclusion}
\vspace{-2mm}


Security and privacy are paramount considerations during electronic communication. Unfortunately, we cannot implement well-known security solutions from computer networks on resource constrained SoCs in embedded systems and IoT devices. Specifically, these security solutions can lead to unacceptable performance overhead. In this paper, we proposed a lightweight encryption and anonymous routing protocol that addresses the classical trade-off between security and performance. Our approach uses a secret sharing based mechanism to securely transfer data in an NoC based SoC. Packets are changed at each hop and the complete packet is constructed only at the destination. Therefore, an eavesdropper along the routing path is unable to recover the plaintext of the intended message. Data is secured using only a few addition and multiplication operations which allows us to eliminate complex cryptographic operations that cause significant performance overhead. Our anonymous routing protocol achieves superior performance compared to traditional anonymous routing methods such as onion routing by eliminating the need for per-hop decryption. 
Experimental results demonstrated that implementation of existing security solutions on NoC can introduce significant (1.5X) performance degradation, whereas our approach can provide the desired security requirements with minor (4\%) impact on performance. 

\section*{Acknowledgments}
This work was partially supported by the National Science Foundation (NSF) grant SaTC-1936040.


\ifCLASSOPTIONcaptionsoff
  \newpage
\fi



%

\bibliographystyle{IEEEtran}
\bibliography{IEEEabrv,bare_jrnl}

%

\newpage

\begin{IEEEbiography}[{\includegraphics[width=1in,height=1.25in,clip,keepaspectratio]{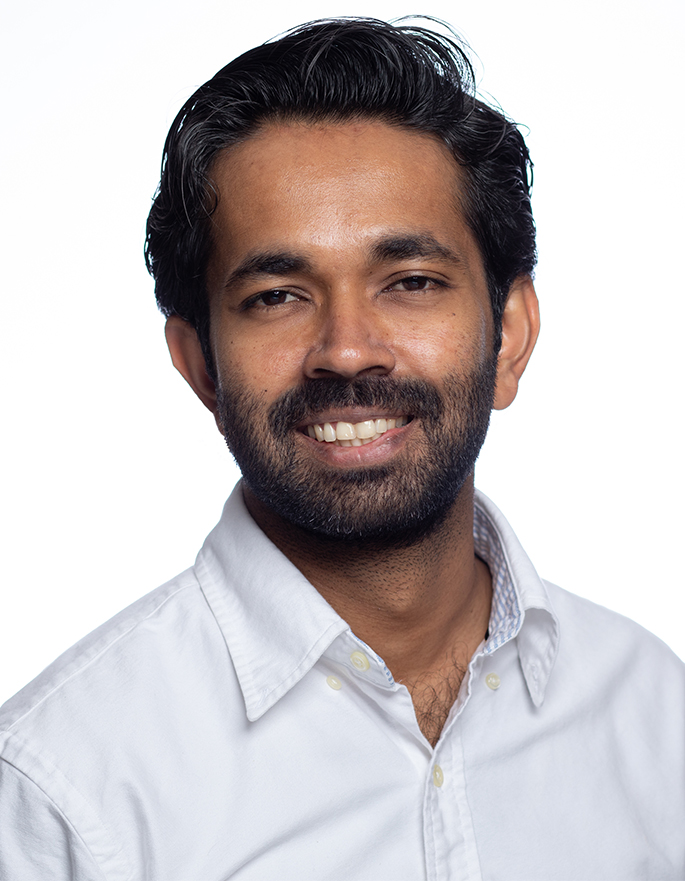}}]{Subodha Charles}
is a Senior Lecturer in the Department of Electronics and Telecommunications Engineering, University of Moratuwa, Sri Lanka. He received his Ph.D in Computer Science from the University of Florida in 2020. His research interests include hardware security and trust, embedded systems and computer architecture.
\end{IEEEbiography}


\begin{IEEEbiography}[{\includegraphics[width=1in,height=1.25in,clip,keepaspectratio]{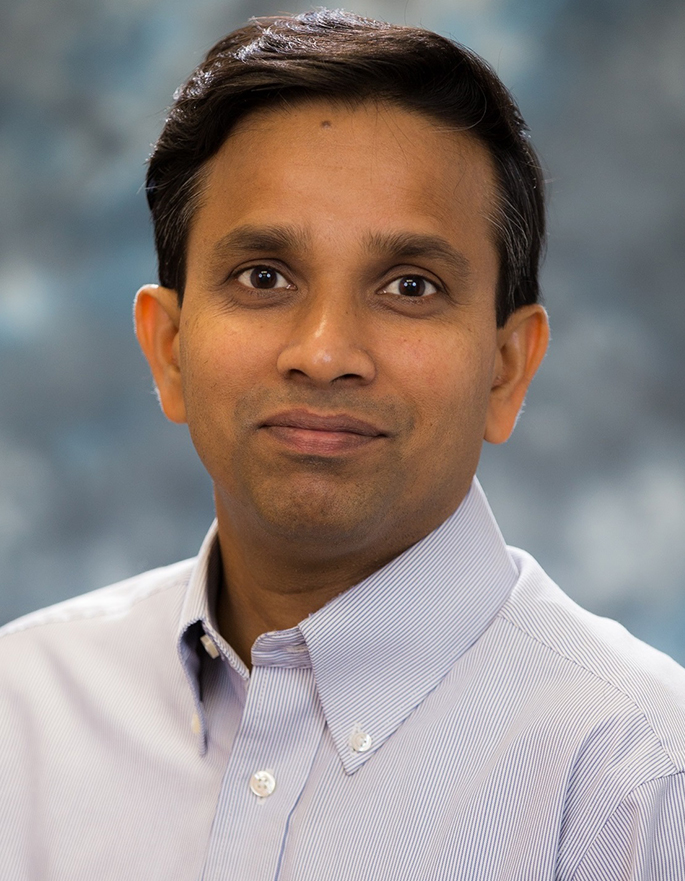}}]{Prabhat Mishra}
is a Professor in the Department of Computer and Information Science and Engineering at the University of Florida. He received his Ph.D. in Computer Science from the University of California at Irvine in 2004. His research interests include embedded and cyber-physical systems, hardware security and trust, computer architecture, energy-aware computing, formal verification, system-on-chip validation, machine learning, and quantum computing. He currently serves as an Associate Editor of ACM Transactions on Embedded Computing Systems and IEEE Transactions on VLSI Systems. He is an IEEE Fellow and ACM Distinguished Scientist.
\end{IEEEbiography}

\vfill




\end{document}